\documentclass[floatfix,aps,twocolumn,showpacs,preprintnumbers,footinbib,longbibliography,prl,superscriptaddress]{revtex4-1}

\usepackage{mathptmx}
\usepackage[autostyle=true]{csquotes}
\usepackage{mathtools}
\usepackage{textcomp}
\usepackage{gensymb}
\usepackage{graphicx}
\usepackage{siunitx}
\usepackage{ulem}
\usepackage{color}
\usepackage[unicode=true,
 bookmarks=true,bookmarksnumbered=true,bookmarksopen=true,bookmarksopenlevel=2,
 breaklinks=false,pdfborder={0 0 1},backref=false,colorlinks=true]
 {hyperref}

 \hypersetup{linkcolor=blue, citecolor=blue, urlcolor=blue, filecolor=blue, pdfpagelayout=OneColumn,
  pdfnewwindow=true, pdfstartview=XYZ, plainpages=false}

\usepackage[all]{hypcap}

\begin{document}


\title{Probing quantum effects with classical stochastic analogs}

\author{R\'emi Goerlich}
\affiliation{ Universit\'e de Strasbourg, CNRS, Institut de Physique et Chimie des Mat\'eriaux de Strasbourg, UMR 7504, F-67000 Strasbourg, France}
\affiliation{Universit\'e de Strasbourg, CNRS, Institut de Science et d'Ing\'enierie Supramol\'eculaires, UMR 7006, F-67000 Strasbourg, France}

\author{Giovanni Manfredi}%
\email{giovanni.manfredi@ipcms.unistra.fr}
\affiliation{ Universit\'e de Strasbourg, CNRS, Institut de Physique et Chimie des Mat\'eriaux de Strasbourg, UMR 7504, F-67000 Strasbourg, France}

\author{Paul-Antoine Hervieux}%
\affiliation{ Universit\'e de Strasbourg, CNRS, Institut de Physique et Chimie des Mat\'eriaux de Strasbourg, UMR 7504, F-67000 Strasbourg, France}

\author{Laurent Mertz}%
\affiliation{NYU-ECNU Institute of Mathematical Sciences at NYU Shanghai, Shanghai, 200062, China
}
\author{Cyriaque Genet}%
\email{genet@unistra.fr}
\affiliation{Universit\'e de Strasbourg, CNRS, Institut de Science et d'Ing\'enierie Supramol\'eculaires, UMR 7006, F-67000 Strasbourg, France}

\date{\today}

\begin{abstract}
We propose a method to construct a classical analog of an open quantum system, namely a single quantum particle confined in a potential well and immersed in a thermal bath. The classical analog is made out of a collection of identical wells where classical particles of mass $m$ are trapped. The distribution $n(x,t)$  of the classical  positions is used to reconstruct the quantum Bohm potential $V_{\rm Bohm} = -\frac{\hbar^2}{2 m}  \frac{\Delta \sqrt{n}}{\sqrt{n}}$, which in turn acts on the shape of the potential wells. As a result, the classical particles experience an effective ``quantum" force. This protocol is tested with numerical simulations using single- and double-well potentials, evidencing typical quantum effects such as long-lasting correlations and quantum tunneling. For harmonic confinement, the analogy is implemented experimentally using micron-sized dielectric beads optically trapped by a laser beam.
\end{abstract}

\maketitle

\section{Introduction}
Analogies in physics constitute a powerful tool for the understanding of complex phenomena.
Not only they enable us to apply our knowledge and intuition of a specific domain to a different field, but  also offer the possibility to transfer experimental results from one branch of physics to another.
For instance, table-top experiments have been used  to get  insight into complex -- and experimentally unreachable -- domains such as quantum gravity and black holes, using acoustic \cite{Barcelo2011} or optic \cite{Roger2016} analogs.
Of particular interest here are classical analogs of quantum systems \cite{Dragoman2004}, based on optic \cite{Bouwmeester1995} or hydrodynamic \cite{Couder2005, Pucci2018} experiments.
These analogs rely on the Madelung representation of the wave function and the corresponding ``hydrodynamic" evolution equations for its amplitude and phase, as in the de Broglie-Bohm version of quantum mechanics \cite{Madelung1927, Bohm1952, Takabayasi1954}.

When a quantum particle is immersed in a thermal bath, and taking the limit of vanishing mass, the hydrodynamic model can be cast in the form of a quantum drift-diffusion (QDD) equation \cite{Degond2005, Pinnau2002}, which is often used to describe charge transport in semiconductor devices.
Here, the QDD equation will be the starting point of our quantum-classical analogy. Indeed, the QDD model has the form of a classical Fokker-Planck equation with the addition of an extra  Bohm potential $V_{\rm Bohm} = -\frac{\hbar^2}{2 m}  \frac{\Delta \sqrt{n}}{\sqrt{n}}$, which depends on the position probability distribution $n(x,t)$ of the particles and  carries the information about quantum correlations.
As is well know, any Fokker-Planck equation is equivalent to a  stochastic process described by a Langevin equation.

Our goal here is to use such underlying classical stochastic process to emulate the evolution of a quantum system.
For the present case, the situation is somewhat more complicated, because the Bohm potential depends on the position probability distribution, making the process nonlinear, as the random variable depends on its own probability density.
These types of stochastic processes are known as McKean-Vlasov processes \cite{McKean1966} and have been extensively studied in the past \cite{kolokoltsov_2010}.

\begin{figure}[ht]
	\hspace{-5mm}
	\centering
		\includegraphics[width=0.95\linewidth]{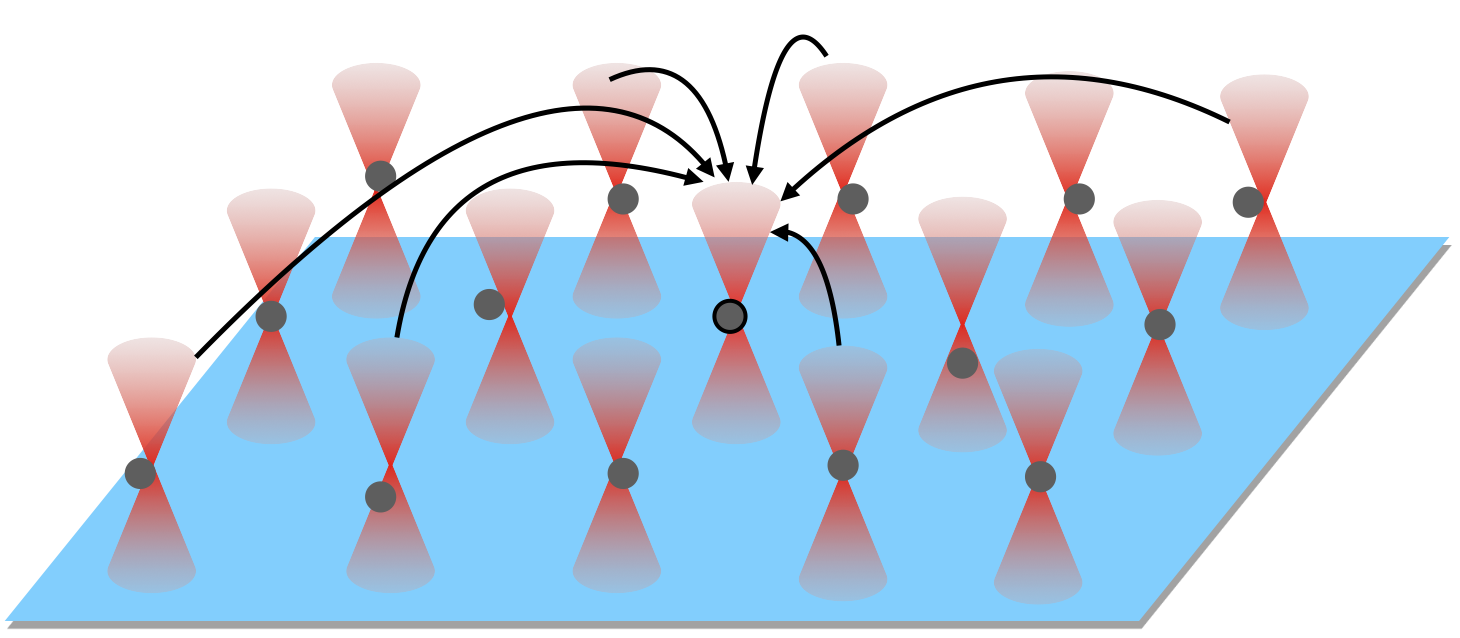}
	\caption{\small{ Schematic view of a possible implementation of the quantum-classical analogs in a multiple optical trapping system. Each identical trap contains a single Brownian particle and the trapping potential, shared among all traps, is controllable. All the particle positions are recorded and the information is collected at each time-step to build the quantum Bohm potential. The latter is then added to the optical trapping potential, thereby acting on all trajectories. This information transfer is represented by the black arrows one chosen trap.
	}}
	\label{fig:schema}
\end{figure}

 Here, we devise a classical analog of this process by reconstructing the probability distribution by statistical means. Our strategy  is based on the possibility of simultaneously manipulating many classical objects, whose ensemble distribution $n(x,t)$ is used as an input to construct the Bohm potential, thus recovering the results of the QDD model.
This can be achieved numerically by simulating $\mathcal{N}$ stochastic trajectories, but, most importantly, can also be realized experimentally, by means of multiple optical trapping of micron-sized Brownian particles \cite{Rosales2020}, as illustrated schematically in Fig. \ref{fig:schema}. Experimentally, up to a few thousand traps can be realized in practice \cite{Bakr2009,wang2020}.

In this work, we will focus on three configurations that nicely capture some typical quantum effects: (i) a quantum increase of the position autocorrelation time, (ii) an analog of the quantum tunneling effect, and (iii) a departure from the classical dynamics for out-of-equilibrium states.
These effects will be investigated with both numerical simulations, (i) and (ii), and optical experiments (iii).

\section{Model}
The dynamics of a quantum particle interacting with a classical thermal environment can be described, in a first approximation, by a Wigner-Boltzmann equation \cite{Manfredi2019} (for a single spatial dimension, which is relevant here):
\begin{equation}
	\begin{aligned}
&\frac{\partial f}{\partial t} + \frac{p}{m}\, \frac{\partial f}{\partial x} -
\frac{i}{2 \pi \hbar^2} \int  e^{\frac{ i (p- p' )   \lambda }{\hbar}} \left[ V_{\rm ext}\left( x_{+} \right) - V_{\rm ext}\left( x_{-}  \right) \right] \times \\
&f ( r,p',t) d{\lambda}\, dp' = Q(f),
	\end{aligned}
\label{wignerequation}
\end{equation}
%
%
where $x_{\pm} \equiv x \pm \lambda /2$, $f(x,p,t)$ is the Wigner phase-space distribution, $V_{\rm ext}(x)$  is the external potential, and $Q(f)$ is a collision operator that models the interaction with the thermal bath. For instance, one could choose a relaxation operator $Q(f)=(f_M-f)/\tau$, where $f_M$ is an equilibrium Maxwellian with bath temperature $T$ and $\tau$ is the thermalization time.
Using a moment expansion of the Wigner-Boltzmann equation \cite{Gardner1994,Manfredi2005,Manfredi2019} and assuming an ideal-gas equation of state for the pressure $P=k_BT n$, one can arrive at a set of two quantum hydrodynamic equations for the density $n(x,t)=\int fdp$ and the mean velocity  $u(x,t) = {1 \over n}\int {p \over m} f dp$:
\begin{eqnarray}
		\frac{\partial n}{\partial t} + \frac{\partial (n u)}{\partial x}&=& 0,  \label{continuity}\\
		\frac{\partial (nu)}{\partial t} + \frac{\partial (nu^2)}{\partial x} &=& -\frac{k_B T}{m } \frac{\partial n}{\partial x} -
\frac{n}{m}  \left(\frac{\partial V_{\rm ext}}{\partial x} + \frac{\partial V_{\rm Bohm}}{\partial x}\right) - \frac{nu}{\tau}, \label{eqmotion}
\end{eqnarray}
where $k_B$ is Boltzmann's constant,  $m$ is the mass of the quantum object, and $\hbar$ is Planck's constant. Quantum effects are contained in the Bohm potential: $V_{\rm Bohm} = -\frac{\hbar^2}{2 m}  \frac{\partial^2_x \sqrt{n}}{\sqrt{n}}$.

It is natural  to choose $\tau = m/\gamma$, where $\gamma$ is the drag coefficient of the object in the fluid that makes up the thermal bath.
Finally, taking the limit $\tau \rightarrow 0$ and $m \rightarrow 0$ , while $m/\tau = \gamma$ remains finite, enables us to drop the inertial terms [left-hand side of Eq. (\ref{eqmotion})] and to inject the expression for $nu$ into the continuity equation \eqref{continuity}, leading to a single quantum drift-diffusion (QDD)  equation for the density \cite{Pinnau2002}:
\begin{equation}
		\frac{\partial n}{\partial t} = \frac{1}{\gamma}\frac{\partial}{\partial x} \left( n \frac{\partial}{\partial x} \left (V_{\rm ext}+ V_{\rm Bohm}\right) \right)+ \frac{k_B T}{\gamma} \frac{\partial^2 n}{\partial x^2} .
	\label{Eq:QDD}
\end{equation}

This equation has the structure of a classical Fokker-Planck equation for Brownian motion, and differs from it only by the  presence of the Bohm potential. This is an important difference, however, as the Bohm potential $V_{\rm Bohm}[n] $ is itself a functional of the probability density $n(x,t)$ and its derivatives. The  stochastic process underlying Eq. (\ref{Eq:QDD}) belongs to the class of McKean-Vlasov processes \cite{McKean1966,kolokoltsov_2010}, describing random variables whose trajectories depend on their own probability distribution.
The Langevin equation of the stochastic process associated with Eq. (\ref{Eq:QDD}) can be written as:
\begin{equation}
dx_t = -\frac{1}{\gamma} \frac{\partial V_{\rm ext} }{\partial x} \,dt -\frac{1}{\gamma} \frac{\partial V_{\rm Bohm}[n]  }{\partial x}\, dt + \sqrt{\frac{2k_BT}{\gamma}}dW_t ,
\label{eq:mckv}
\end{equation}
where $dW_t$ is the Wiener increment due to white noise, with zero mean $\langle dW_t \rangle = 0$ and no memory $\langle dW_t dW_s \rangle = \delta(t-s) dt$.

The stochastic/diffusive model of Eqs.  (\ref{Eq:QDD}) and (\ref{eq:mckv}) describes a quantum system coupled to a classical thermal bath. The coupling will induce some level of decoherence, leading to the partial loss of the quantum character of our system. Nevertheless, some quantum properties will persist in spite of the decoherence, notably those encapsulated in the Bohm potential. The aim of this work is to emulate such quantum properties using a purely classical experimental setup.

In order to emulate Eq. (\ref{eq:mckv}) the key issue is to be able to inject the probability distribution $n$ into the stochastic process itself. This can be achieved with a classical system if one can generate (numerically or experimentally) $\mathcal{N}$ simultaneous trajectories in order to reconstruct $n(x,t)$, and hence the Bohm potential, at each time-step. Experimentally, this may be implemented using a multiple optical trapping system (see Fig. \ref{fig:schema}).

Equation (\ref{eq:mckv}) can be rewritten in a normalized form that brings out a dimensionless parameter $\epsilon$, which plays the role of a normalized Planck constant and governs the strength of the quantum effects.
The quantity $\epsilon^2$ can be interpreted as the ratio between the quantum decoherence time and the classical relaxation time
(see Appendix \ref{APP:adim_procedure}).
In our classical analog, $\epsilon$ is no longer related to Planck's constant, but can be adjusted at will, within the practical limits of the experimental or numerical realization. The classical case, i.e. standard Brownian motion, corresponds to $\epsilon = 0$, while when $\epsilon \approx 1$ ``quantum" effects play a significant role.

\section{Numerical results}
We use a quartic external potential $V_{\rm ext} = \alpha x^2 + \beta x^4$. We consider two cases, with either  $\alpha>0$ (anharmonic single well) or $\alpha<0$ (bistable double-well potential), and focus on the features of the equilibrium distribution. Transients will be analyzed later using an experimental protocol. Simulations are performed with a first-order Euler-Maruyama algorithm \cite{volpe, kloeden} that solves the McKean-Vlasov equation (\ref{eq:mckv})  for $\mathcal{N}$ trajectories $x(t)$ simultaneously.
At each time-step, a smooth distribution $n(x,t)$ is constructed from the $\mathcal{N}$ trajectories by softening the particle positions $x(t)$ with a Gaussian kernel. Details are given in the Appendix \ref{APP:numerical method}.
We take as the initial condition the stationary distribution of a classical process ($\epsilon = 0$), then turn on quantum effects ($\epsilon > 0$) and let the system evolve to its new equilibrium.

\begin{figure}[ht]
	\centering
		\includegraphics[width=0.85\linewidth]{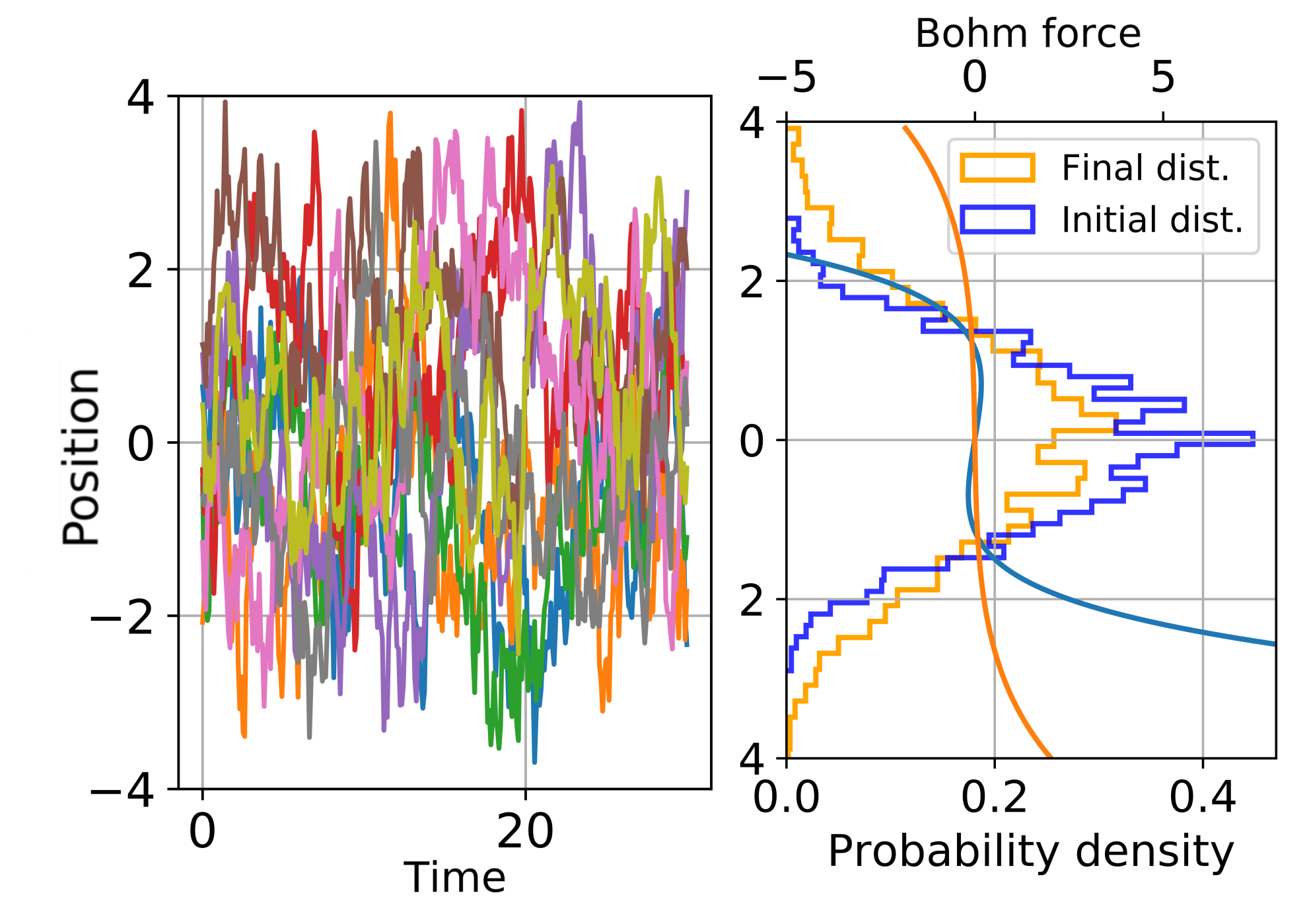}
		\includegraphics[width=0.85\linewidth]{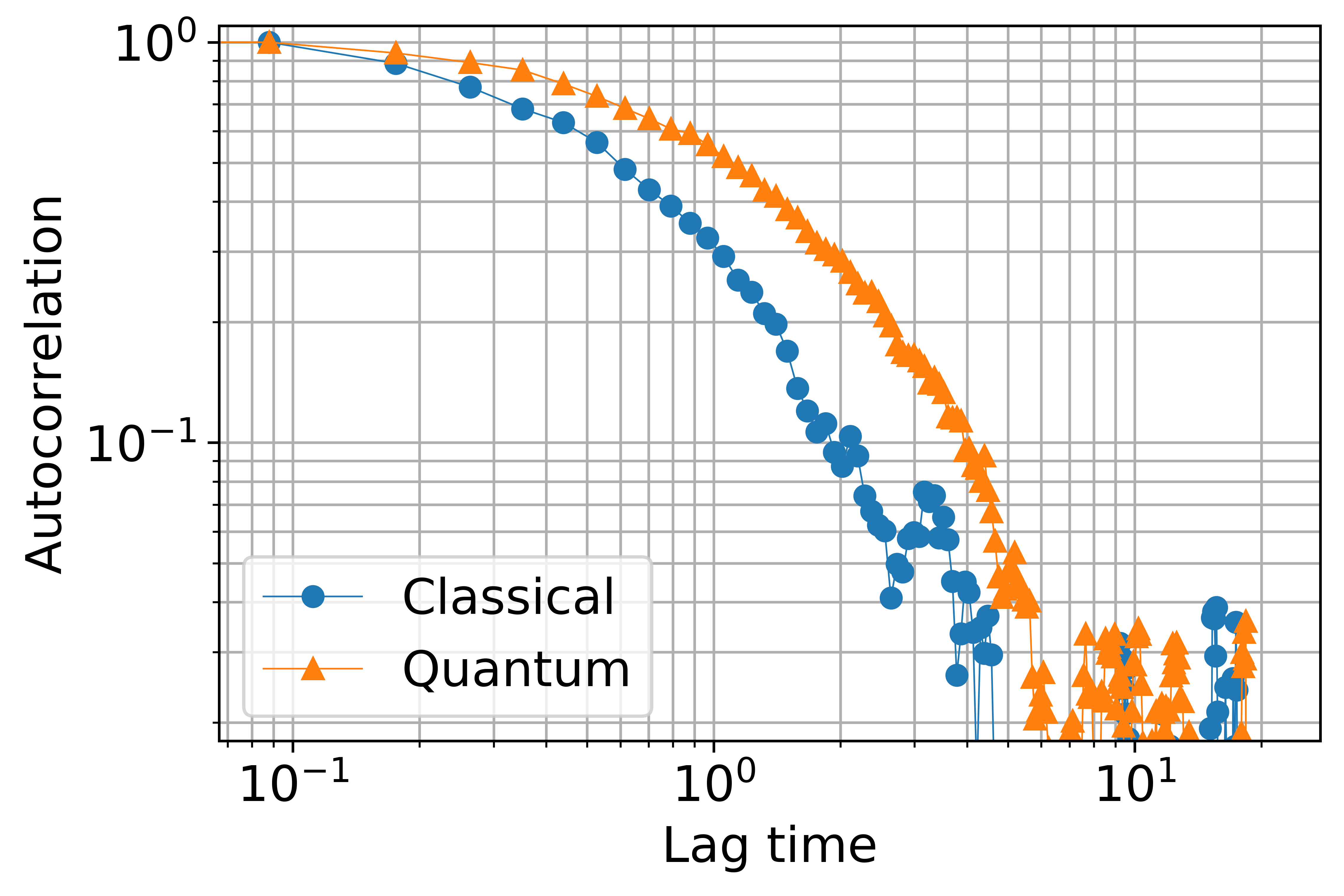}
	\caption{\small{(Top left) Evolution of 100 simulated trajectories, initially distributed according to the classical stationary state in the quartic potential $V_{\rm ext} = \alpha x^2 + \beta x^4$, where $\alpha = 0.6$ and $\beta = 0.2$, computed with a total $\mathcal{N} = 3000$ trajectories with $\epsilon = 4$, for $300$ time-steps with $dt = 10^{-1}$; (top right) Histograms of the initial (classical, $\epsilon=0$) and final (quantum, $\epsilon=4$) equilibrium distributions, together with the average Bohm force (solid line); (bottom) Logarithmic plot of the normalized ensemble-averaged correlations $\langle x(\tau)  x(t_0) \rangle/\langle x^2(t_0) \rangle$ as a function of the lag time $\tau$, for trajectories $x(t)$ undergoing a quantum McKean-Vlasov  (orange triangles) or  classical (blue circles) stochastic process.}}
	\label{fig:TrajDist}
\end{figure}
\vspace{3mm}

In the $\alpha > 0$ case, the confining potential is a single quartic well. In Fig. \ref{fig:TrajDist} (top left) we represent some simulated trajectories. After a certain relaxation time, the system stabilizes around a quantum equilibrium distribution, which differs significantly from the initial classical Boltzmann equilibrium $n \propto e^{-V_{\rm ext}/k_B T}$ (top right).
This departure from the classical result is due to the Bohm force, which works against the external confinement, as detailed in the Appendix \ref{APP:distrib}.

In Fig. \ref{fig:TrajDist} (bottom frame), we show the normalized autocorrelation at equilibrium: $\langle x(\tau)  x(t_0) \rangle/\langle x^2(t_0) \rangle$, where the average is over all the trajectories, as a function of the lag-time $\tau$.
The initial time is set at an instant $t_0$, when the distribution has already relaxed to its quantum equilibrium. We note that the addition of the quantum Bohm potential induces longer-lasting correlations compared to the classical case. A straightforward interpretation is that the McKean-Vlasov trajectories are correlated with one another through the action of the Bohm force.

\begin{figure}[ht]
		\includegraphics[width=0.85\linewidth]{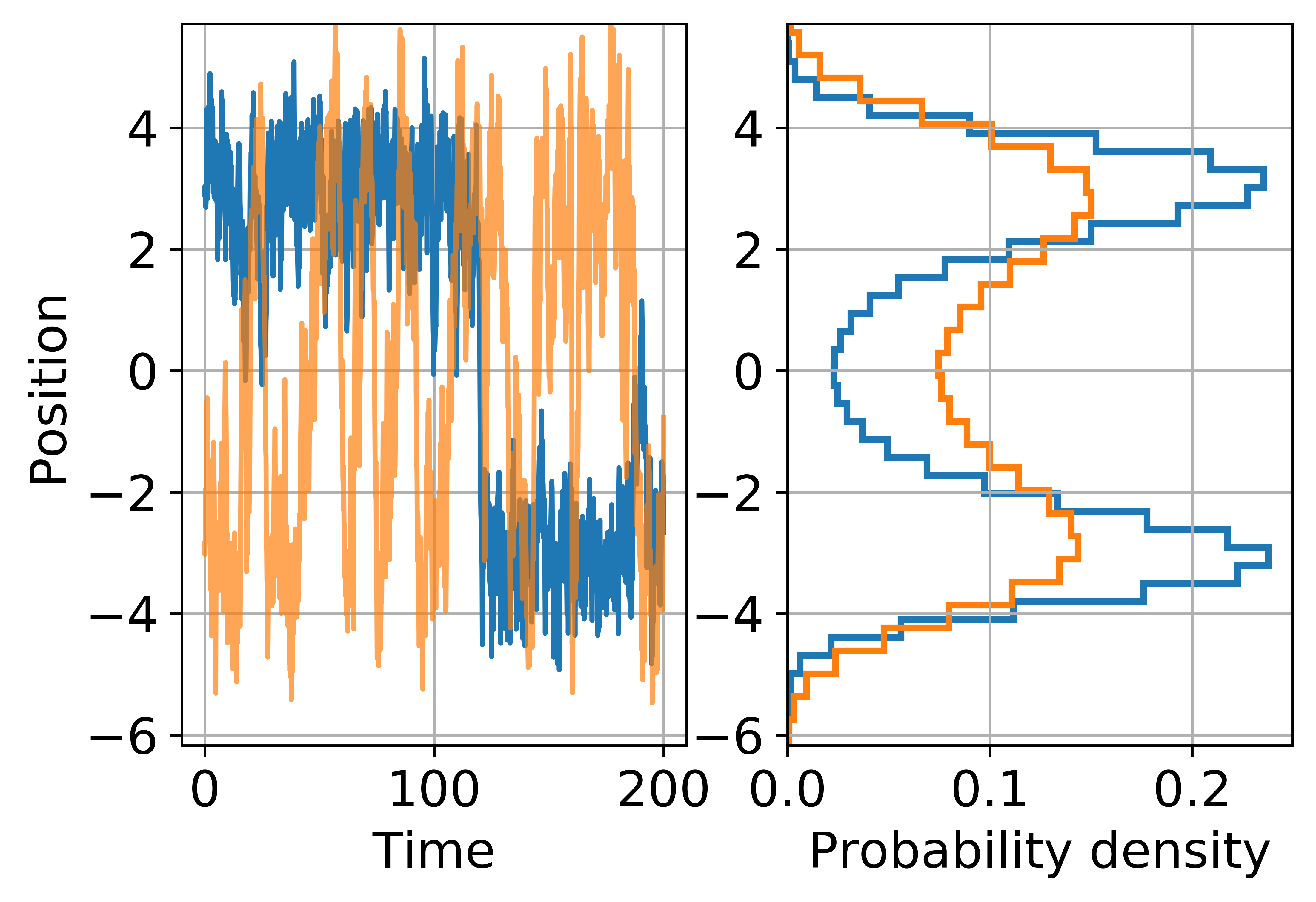}
		\includegraphics[width=0.85\linewidth]{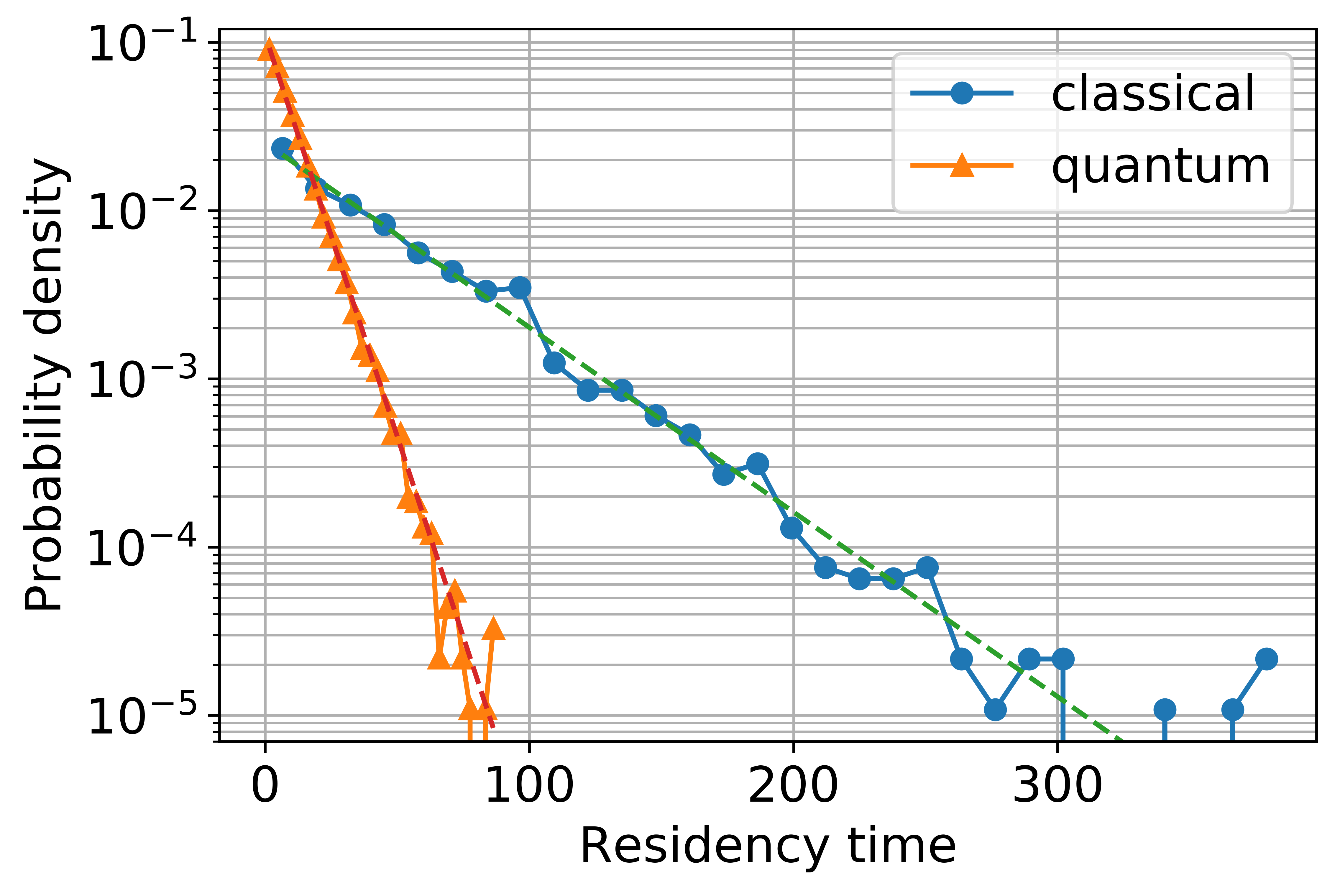}
	\caption{\small{(Top left) Classical ($\epsilon=0$, blue) and McKean-Vlasov ($\epsilon=2$, orange) trajectories in a  double-well potential ($\alpha = - 1$, $\beta = 0.1$) and (top right)  their respective probability distributions, for $\mathcal{N} = 3000$ trajectories simulated for $2000$ time-steps with $dt = 10^{-1}$;
(bottom) Probability distribution of the residency times $\tau_R$ for the classical (blue circles) and quantum (orange triangles) cases. The straight lines represent the corresponding Poisson distributions, with $ \langle \tau_R \rangle = 42.9$ for the classical and $\langle \tau_R \rangle = 8.4$ for the McKean-Vlasov case.}}
\label{fig:Duffing}
\end{figure}

We now turn to the case $\alpha < 0$, for which the confining potential is a bistable double well.
Using the same numerical method, we simulate $\mathcal{N} =3000$ trajectories for both the classical ($\epsilon=0$) and the McKean-Vlasov ($\epsilon = 2$) stochastic processes.
In the classical case, the trajectories linger in one of the wells for a relatively long residency time $\tau_R$, before occasionally jumping to the second well due to thermal fluctuations. In contrast, in the McKean-Vlasov case these jumps occur much more frequently  (Fig. \ref{fig:Duffing}, top left frame).
The jump events are correctly described by Poisson statistics \cite{Grebenkov2014, Mandel79} and the probability distribution of the residency times obeys an exponential decay law \cite{Libchaber1992} $P(\tau_R) = \lambda e^{-\lambda \tau_R} $, where $\lambda \equiv 1/\langle \tau_R \rangle$. The results shown in Fig. \ref{fig:Duffing} (bottom frame) are in good agreement with this Poissonian law, both for the classical and for the McKean-Vlasov processes, albeit with different values of $\lambda$,  the effect of the Bohm potential being to decrease the mean residency time.
The enhanced mobility between the two wells  is clearly seen in the probability distribution of the particle positions (top right frame), which signals a decrease of the effective potential barrier due to the quantum  Bohm potential. This result can be interpreted as a manifestation of quantum tunneling, which increases the frequency of barrier-crossing events beyond the classical thermally-induced probability.
We emphasize that this quantum-like property persists despite the decoherence that is inherent to our model.

\section{Experimental realization}
Next, we turn to the possibility of implementing experimentally our classical-quantum analog. For this, we adopt a harmonic confinement potential $V_{\rm ext} = \kappa x^2 / 2$, which is easy to realize with an optical trap, and also allows us to
circumvent the need of using many traps to implement the analog process. This approach will be used to study the effect of the Bohm potential in an out-of-equilibrium configuration.

First, we note \cite{Burghardt2002} that  a Gaussian probability distribution: $ n(x,t) = \left[2 \pi S(t)\right]^{-1/2} e^{-x^2/2S(t)} $, where $S(t)$ is the time-dependent variance of the distribution, is an exact solution of the McKean-Vlasov process (\ref{eq:mckv}), provided the variance obeys the following equation:
\begin{equation}
\frac{dS(t)}{dt} = - \frac{2\kappa}{\gamma} S(t) + \frac{\hbar^2}{2 m \gamma S(t)}  + \frac{2k_B T}{\gamma} .
\label{Eq:VarGaussian}
\end{equation}
Furthermore, for such Gaussian distribution the Bohm force takes a simple analytical form: $ \partial_x V_{\rm Bohm}(x, t) = \frac{\hbar^2}{4m} \frac{x_t}{S^2(t)}$. In this case, both the external force and the Bohm force have the same functional form, linear in the stochastic variable $x_t$, and can therefore be grouped together into a single harmonic term with modified stiffness: $\bar{\kappa}(t) \equiv \kappa(t) - \frac{\hbar^2}{4m S^2(t)} $. Hence, the quantum McKean-Vlasov process can be expressed as an ordinary (Ornstein-Uhlenbeck) stochastic process:
\begin{equation}
dx_t = \frac{-\bar{\kappa}(t)}{\gamma} x_t dt + \sqrt{\frac{2k_BT}{\gamma}}dW_t.
\label{Eq:Mck_V_gaussian}
\end{equation}

Despite this apparent mathematical simplicity, all the physical richness of the analog model is preserved, with the modified stiffness $\bar{\kappa}(t)$ still depending on the ensemble variance $S(t)$ as a consequence of the quantum nature of the problem.
Moreover, in this harmonic case, the dimensionless parameter governing quantum effects takes the form: $\epsilon^2 \equiv \frac{\hbar^2 \kappa}{2m\, (k_BT)^2} = \lambda^2_{\rm dB}/2\lambda^2_{\kappa}$, i.e. half the ratio between the de Broglie thermal wavelength $\lambda_{\rm dB}=\hbar/\sqrt{m k_B T}$ and the classical width of the harmonic oscillator at thermal equilibrium $\lambda_{\kappa}=\sqrt{ k_B T/\kappa}$.
The specificity of the harmonic confinement is that the variance need not be measured out of a collection of  trajectories taking place simultaneously in $\mathcal{N}$ identical traps, as in Fig. \ref{fig:schema}. Instead, $S(t)$ can be computed from Eq. (\ref{Eq:VarGaussian}) and then used to construct the Bohm potential or force, thus avoiding the necessity of using many optical traps in the experiment.
\begin{figure}[htb]
	\hspace{-7mm}
	\centering{
		\includegraphics[width=0.9\linewidth]{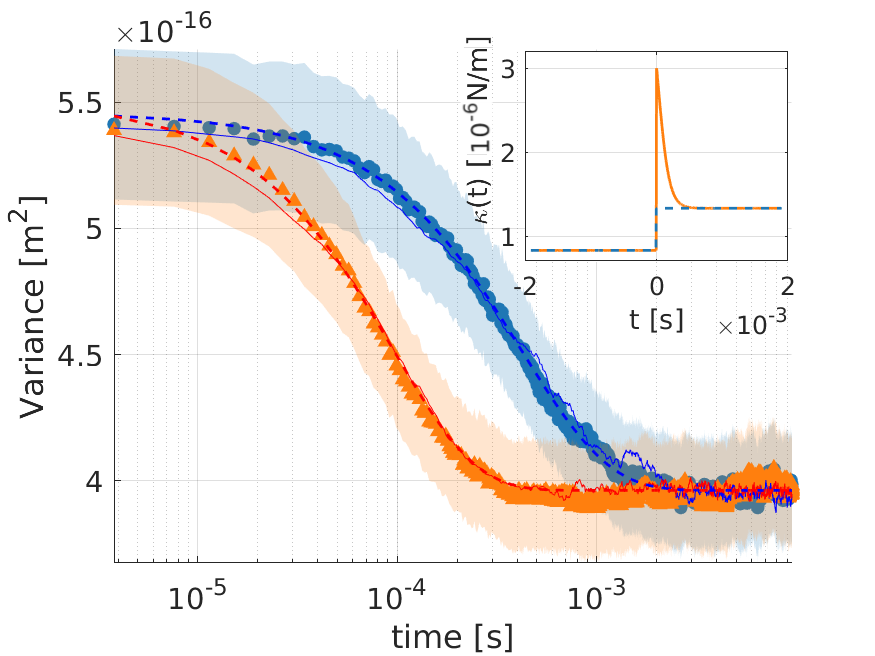}}
	\caption{\small{Experimental variance of an ensemble of over $\mathcal{N} = 2 \times 10^4$ trajectories during the transient between two harmonic wells with different stiffness. The parameter governing the strength of the Bohm force is $\epsilon = 1.8$. We show results of the quantum McKean-Vlasov  (orange triangles) and classical Ornstein-Uhlenbeck (blue circles) dynamics. The colored patches corresponds to a $99.7 \%$ confidence interval, taking into account both experimental and statistical errors -- see Appendix \ref{APP:setup}. On each curve, we superimpose the result of a numerical simulation performed by measuring at each instant the ensemble variance of $\mathcal{N}$ simultaneous trajectories and using it to compute the Bohm force (respectively, red and blue thin solid lines). We also show the result of Eq. (\ref{Eq:VarGaussian}) for the variance (blue and red dashed lines).  The inset shows the evolution of the stiffness $\bar{\kappa}(t)$ (orange line) as well as the equivalent classical step $\bar{\kappa}_{cl}(t)$ (blue dashed line).
	 }}
	\label{fig:ExpResults}
\end{figure}
\vspace{3mm}

Our experimental setup -- presented in detail in the Appendix \ref{APP:setup} -- is composed of a single $1 \si{\micro \meter}$ dielectric bead optically trapped by a $785 \si{\nano\meter}$  Gaussian laser beam.
The optical potential created by the gradient forces at the waist of the beam is harmonic, with a stiffness that is proportional to the intensity of the laser and  can thus be controlled precisely.
The bead is immersed in water at ambient temperature and undergoes Brownian motion due to the thermal fluctuations.
The overall motion is consistent with an Ornstein-Uhlenbeck process and is therefore suited to implement our model.

Here, we  use the above approach to study out-of-equilibrium evolutions with a time-dependent stiffness $\kappa(t)$.
The simplest possible out-of-equilibrium process is the transient occurring when the stiffness $\kappa(t)$ is suddenly changed from an initial value $\kappa_i$ to a final value $\kappa_f$ (step-like protocol). The system is at thermal equilibrium at the initial and final times. The transient evolution of the variance can be  computed using Eq. (\ref{Eq:VarGaussian}), allowing us to construct the modified stiffness $\bar{\kappa}(t)$, which evolves from $\bar{\kappa}_i$ to $\bar{\kappa}_f$ in a non-trivial way due to the influence of $S(t)$.
One can argue that, since different values of $\epsilon$  lead to different values of $\bar{\kappa}$ for the initial and final equilibria,  the classical ($\epsilon=0$) and quantum (here, $\epsilon = 1.8$) transients are difficult to compare, as they do not begin and end with the same values of the stiffness. With this in mind, we also implemented an equivalent classical protocol $\bar{\kappa}_{\rm cl}(t)$ that goes from $\bar{\kappa}_i$ to $\bar{\kappa}_f$ in a step-like way, i.e. without the dynamical influence of the Bohm force.
These two protocols, represented in the inset of Fig. \ref{fig:ExpResults}, connect the same initial and final equilibria, and are thus well-suited to compare the classical and quantum dynamics out-of-equilibrium.

Finally, in order to obtain ensemble averages out of our single trajectory, we rely on the ergodic hypothesis and use a time-series of trajectories instead of a statistical ensemble. We send the same $\kappa(t)$ protocol at a low enough repetition rate so that equilibrium is reached between two consecutive events, and then reconstruct a synchronized ensemble from this time series.
The result is an ensemble of over $\mathcal{N} = 2 \times 10^4$  trajectories experiencing a given protocol, either $\bar{\kappa}(t)$ in the quantum case or $\bar{\kappa}_{\rm cl}(t)$ in the classical case.

The main observable of interest here is the time evolution of the ensemble variance, represented in Fig. \ref{fig:ExpResults} for both the quantum and classical cases. Our measurements clearly reveal the influence of the Bohm force on $S(t)$. Strikingly, the addition of an effective quantum force accelerates its relaxation, and this for all selected values of $\epsilon$, as detailed in the Appendix \ref{APP:setup}. Looking at $\bar{\kappa}(t)$ in the McKean-Vlasov process (Fig. \ref{fig:ExpResults}, inset), the acceleration appears as the result of a strong and sudden reduction of the optical trapping volume under the influence of the quantum Bohm force field.
On each curve, we also represent the result of numerical simulations, where the evaluation of the Bohm term is not performed through the solution of Eq. (\ref{Eq:VarGaussian}), but by actually computing the ensemble variance of $\mathcal{N} = 2 \times 10^4$ distinct trajectories at each time-step.
The agreement of both the experimental and numerical results with the analytical solution of Eq. (\ref{Eq:VarGaussian}) is quite remarkable.

\section{Conclusion}
We highlighted an analogy between an open quantum system immersed in a thermal bath and a classical nonlinear  stochastic process (McKean-Vlasov process).
This correspondence opened up the possibility to build a classical analog of the quantum model, by evolving  many stochastic trajectories in parallel and using their distribution to reconstruct the quantum Bohm potential.
This classical analog was realized both numerically and experimentally, evidencing typical quantum effects such as long-lasting correlations and quantum tunneling.

The present work is a first step in the experimental implementation of classical analogs of quantum systems using optically trapped Brownian particles, which should provide an original  platform  for probing the subtle connections between quantum mechanics and stochastic thermodynamics. We emphasize that our approach is obviously not limited to the QDD framework,  whose validity is constrained by several conditions. It can, for instance, be extended to classically emulate fully quantum evolutions, described in the most general scenario by the time-dependent Schr{\"o}dinger equation. This non-trivial generalization should lead to classical stochastic systems specifically designed to access the controlled exploration of many fundamental quantum effects, such as quantum correlations, quantum decoherence, and quantum chaos.

\medskip
\noindent
{\bf Acknowledgements}\\
This work of the Interdisciplinary Thematic Institute QMat, as part of the ITI 2021 2028 program of the University of Strasbourg, CNRS and Inserm, was supported by IdEx Unistra (ANR 10 IDEX 0002), by SFRI STRAT'US project (ANR 20 SFRI 0012), by the Labex NIE (ANR-11-LABX-0058 NIE) and CSC (ANR-10-LABX-0026 CSC) projects, and by the University of Strasbourg Institute for Advanced Study (USIAS) (ANR-10-IDEX-0002-02) under the framework of the French Investments for the Future Program.
%


\appendix

\section{Nondimensionalization of the McKean-Vlasov process}
\label{APP:adim_procedure}

In order to derive a non-dimensional description of the QDD equation for $n(x,t)$, we start with

\begin{equation}
\frac{\partial n}{\partial t}  = \frac{k_B T}{\gamma}  \Delta n - \frac{\hbar^2}{2 m \gamma} \nabla \cdot \left[n \nabla \left( \frac{\Delta \sqrt{n}}{\sqrt{n}}\right) \right] + \frac{1}{\gamma} \nabla \cdot \left( n \nabla V \right),
\end{equation}

and make the following change of variables, focusing on a quadratic external potential $V_{ext} = \frac{1}{2} \kappa(t) x^2$ for simplicity:

\begin{equation}
\left\{
\begin{aligned}
t &\rightarrow \tilde{t} = t/\tau_{relax} ,\\
x &\rightarrow \tilde{x} = x/\lambda_{\kappa} ,\\
V = \frac{1}{2}\kappa x^2 &\rightarrow \tilde{V} = \frac{1}{2}\kappa x^2 / \kappa_i \lambda_{\kappa}^2 .		
\end{aligned}
\right.
\end{equation}

Here, $\kappa_i$ corresponds to the initial stiffness, $ \tau_{relax} = \gamma / \kappa_i $ is the corresponding relaxation time and $\lambda_{\kappa} \equiv \sqrt{\frac{k_BT}{\kappa_i}}$ the classical width of the harmonic oscillator at thermal equilibrium fixed by equipartition. This change of variables leads to

\begin{equation}
\frac{\partial n}{\partial \tilde{t}}  = \Delta n - \epsilon^2 \nabla \cdot \left[n \nabla \left( \frac{\Delta \sqrt{n}}{\sqrt{n}}\right) \right] + \nabla \left[ n \tilde{\kappa}(t) \tilde{x} \right]
\end{equation}

where $\epsilon^2 \equiv \frac{\hbar^2\kappa_i}{2 m (k_BT)^2}$ is the dimensionless parameter described in the main text.
The non-dimensional stochastic McKean-Vlasov process then writes as:
\begin{equation}
d\tilde{x_t} = - \nabla \tilde{V}_{ext} d\tilde{t} + \epsilon^2 \nabla \left( \frac{\Delta \sqrt{n}}{\sqrt{n}}\right) d\tilde{t} + \sqrt{2}d\tilde{W_t}.
\label{eq:AdimMckv}
\end{equation}

As discussed in the main text, $\epsilon^2 $ can be written as half the ratio between the de Broglie thermal length $\lambda_{\rm dB}$ and the classical width of the harmonic oscillator at thermal equilibrium $\lambda_{\kappa}$.
Another possible interpretation can be given as the ratio between the quantum decoherence time and the thermal relaxation time $\tau_{relax}$.
Following \cite{Zurek2007}, the loss of quantum coherence is governed by a typical time scale  $\tau_D = \tau_{relax} \left(\frac{\hbar^2}{2 m k_B T \Delta x^2}\right)$ where $\Delta x$ is a typical length scale of motion. In our case, we take $\Delta x = \lambda_{\kappa} = \sqrt{k_BT/\kappa} $.
This gives $\epsilon^2 = \tau_D / \tau_{relax}$, i.e., the ratio between the decoherence time and the relaxation time.

\section{Numerical method}
\label{APP:numerical method}

Our numerical approach is based on the Euler-Maruyama stochastic algorithm \cite{volpe, kloeden}.
It consists in a discrete approximation of the stochastic differential equation up to order $\mathcal{O}(\delta t^{1/2})$ in the time increment $\delta t$.
For the stochastic equation
\begin{equation*}
dx_t = - \nabla V_{ext} dt + \nabla \left( \frac{\Delta \sqrt{n}}{\sqrt{n}}\right) dt + \sqrt{2}dW_t,
\end{equation*}
where all variables are non-dimensional, it takes the form
\begin{equation}
	y_{i+1} = y_i -\nabla V_{ext}(y_i) \Delta t + \nabla \left( \frac{\Delta \sqrt{n(y_i)}}{\sqrt{n(y_i)}}\right) \Delta t + \sqrt{2\delta t} \xi_i
\end{equation}
where $y_i$ is the numerical approximation of $x(t)$, $\xi_i$ is a normally distributed random variable, and $\delta t$ is the time-step.

Our numerical approach is the following: like for a classical stochastic algorithm, we compute the positions at a certain time-step for $\mathcal{N}$ Brownian particles simultaneously and then evaluate the forces for the next time-step.
The specificity of the present case arises from the evaluation of the quantum force, which requires the reconstruction of the density $n(y_i)$.
We use a particle-in-cell method to evaluate $n$: to each particle we associate a Gaussian kernel of width $\sigma$, and the sum of such Gaussians will yield a reasonably smooth estimation of the density $n$.
From this reconstructed density $n(y_i)$ we are able to compute the quantum force $\nabla \left( \frac{\Delta \sqrt{n(y_i)}}{\sqrt{n(y_i)}}\right)$, which in turn is used to update the positions of the particles at the next time-step.

\begin{figure}[ht]
	\centering
	\includegraphics[width=0.8\linewidth]{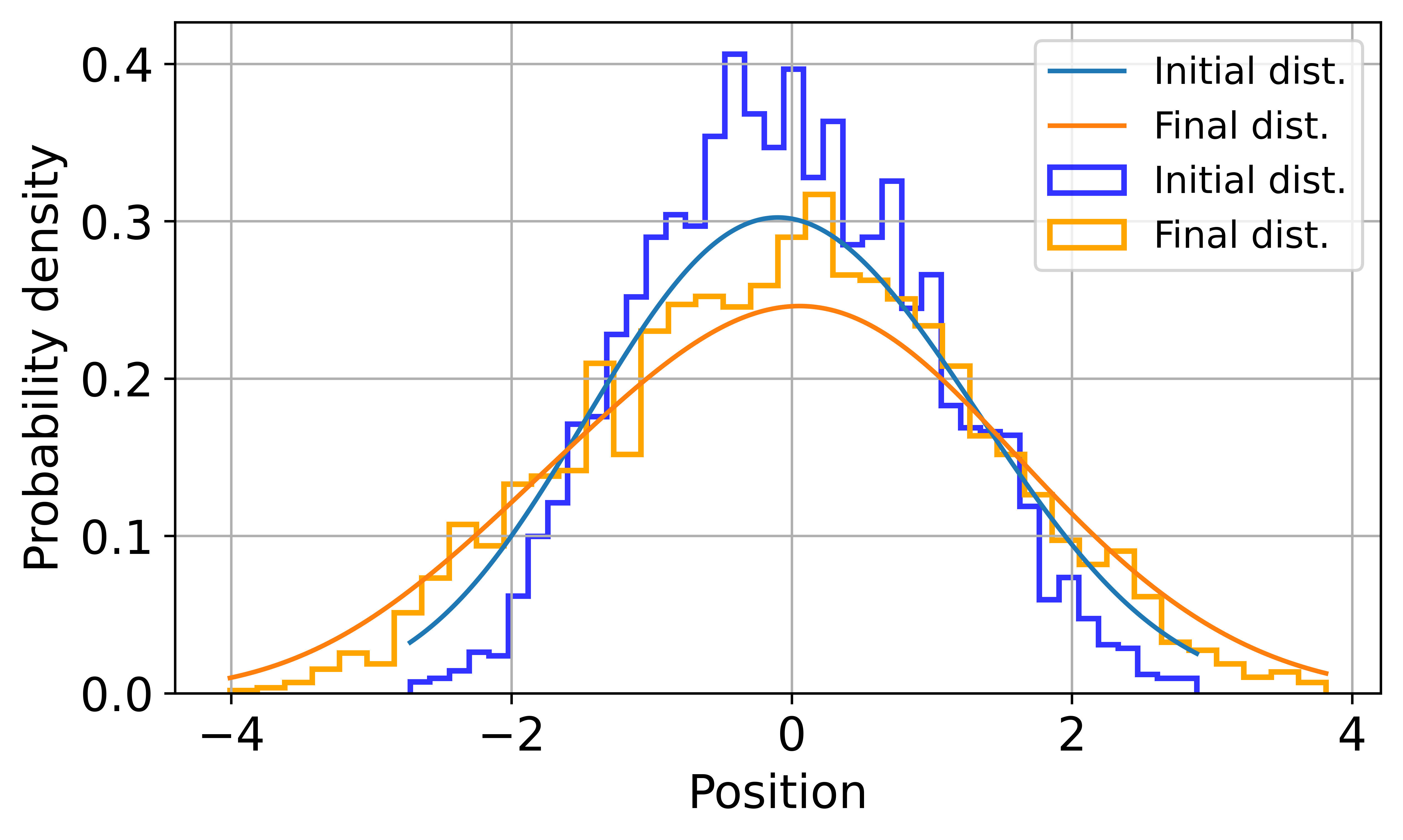}
	\caption{\small{
			Histograms of the particles distributions and corresponding smooth densities obtained as the sum of Gaussian kernels for $3000$ trajectories. We note that the choice of the kernel wifth leads to a flattening of the distribution.
	}}
	\label{fig:DistFitKDE}
\end{figure}

One important parameter here is the width $\sigma$ of the Gaussian kernel. Too small a width will lead to a noisy evaluation of the density, which is dangerous because the quantum force requires the computation of the third derivative of the density.
The problem mainly arises when we use few trajectories, as we do in our work in order to stay close to the experiments.
In our case, where $3000$ trajectories are employed, we made a reasonable compromise between a kernel of width $\sigma \approx 0$, i.e., a sum of Dirac delta functions, and a width $\sigma \approx 1$ (in normalized units) which would amount to a Gaussian approximation for the whole density.
We use a value  $\sigma=0.8$ which is large enough to smooth out irregularities arising from the small statistical sample, but small enough to account for departures from a perfectly Gaussian density (as evidenced from the skewness and kurtosis, see Appendix \ref{APP:distrib}).
The histograms of the particles distributions and the corresponding smooth densities are shown in Fig. \ref{fig:DistFitKDE}.

\vspace{5mm}

\section{Statistical analysis}
\label{APP:distrib}
\begin{figure}[ht]
			\includegraphics[width=0.7\linewidth]{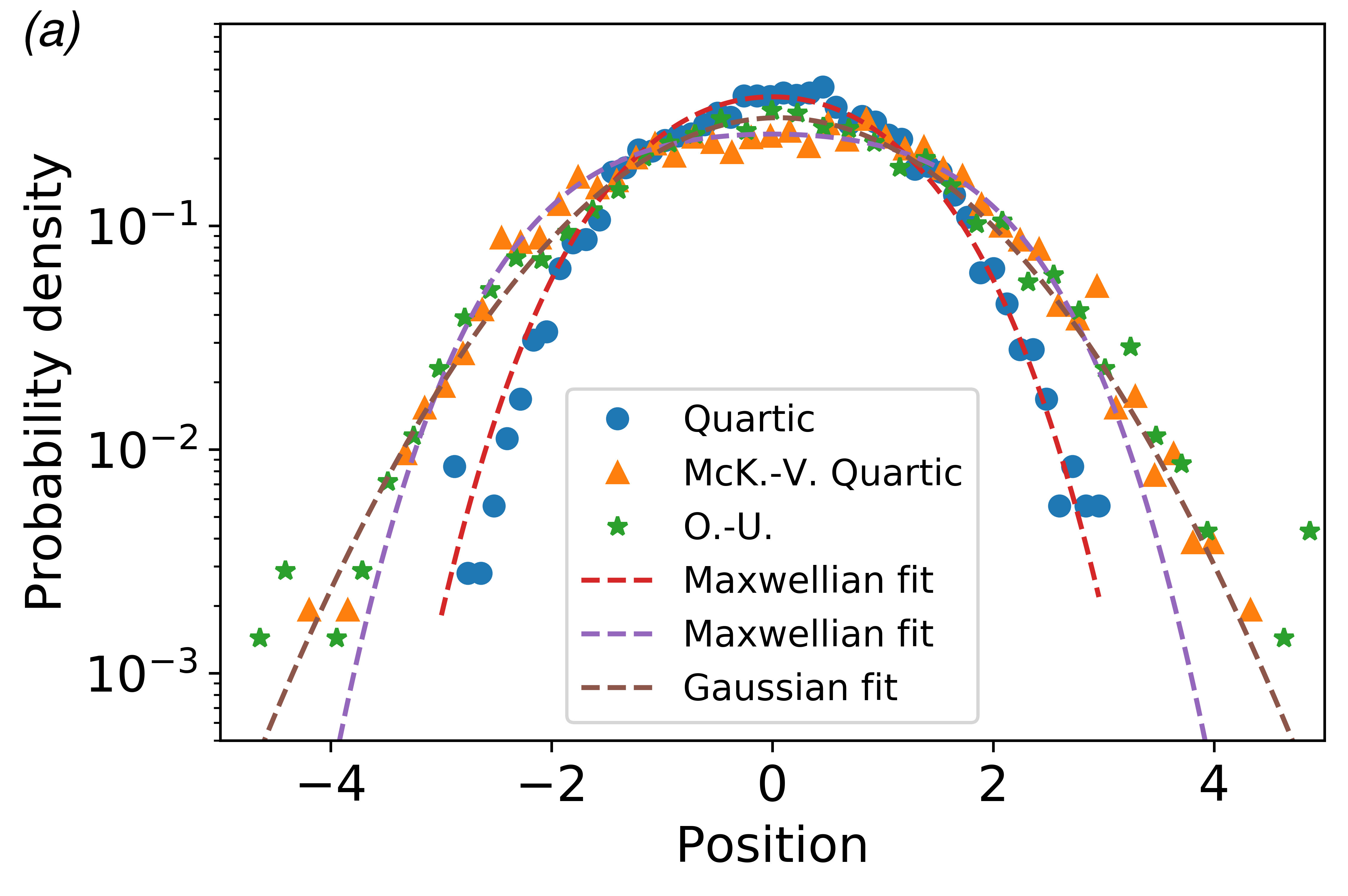}
			\includegraphics[width=0.7\linewidth]{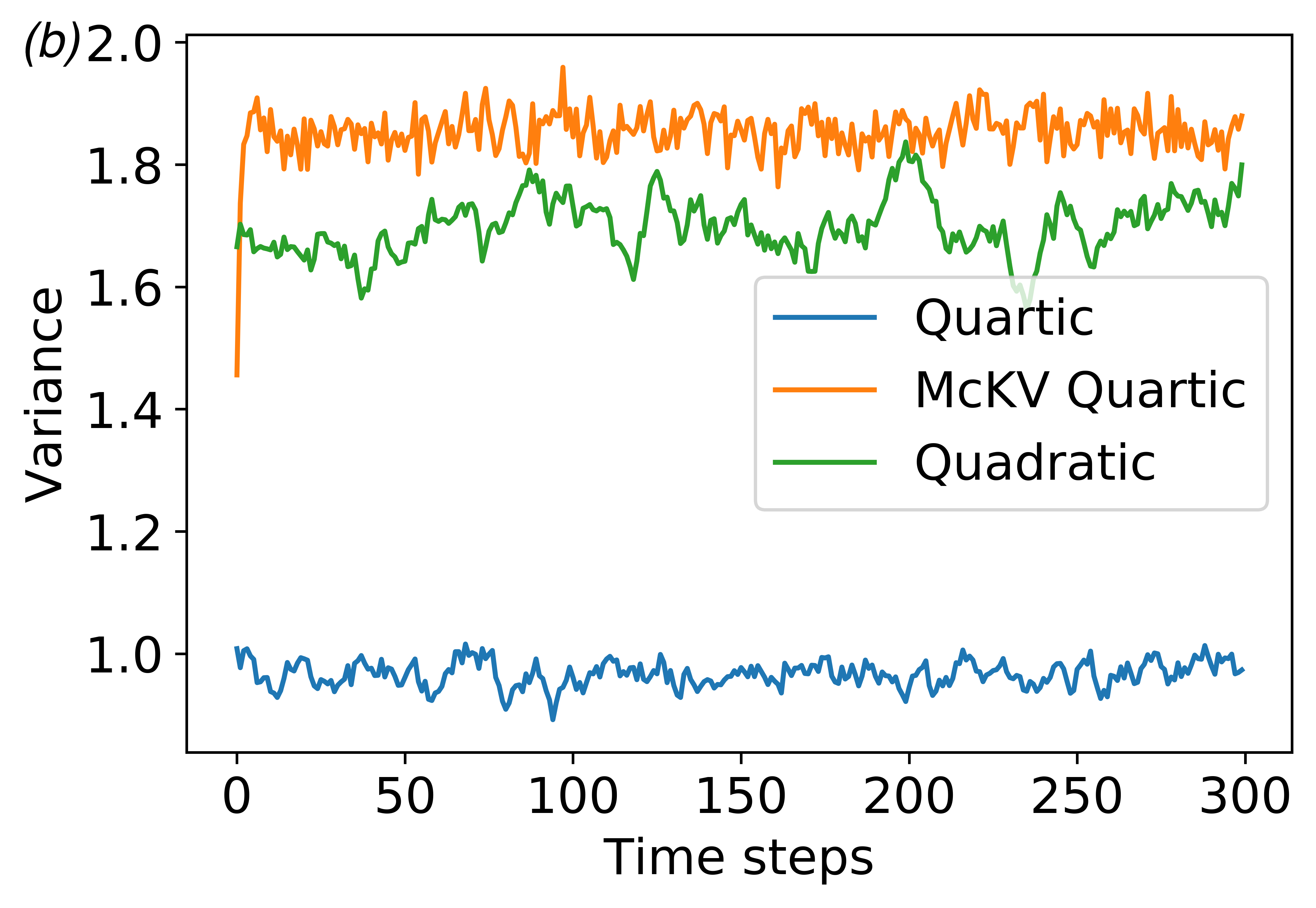}
			\includegraphics[width=0.7\linewidth]{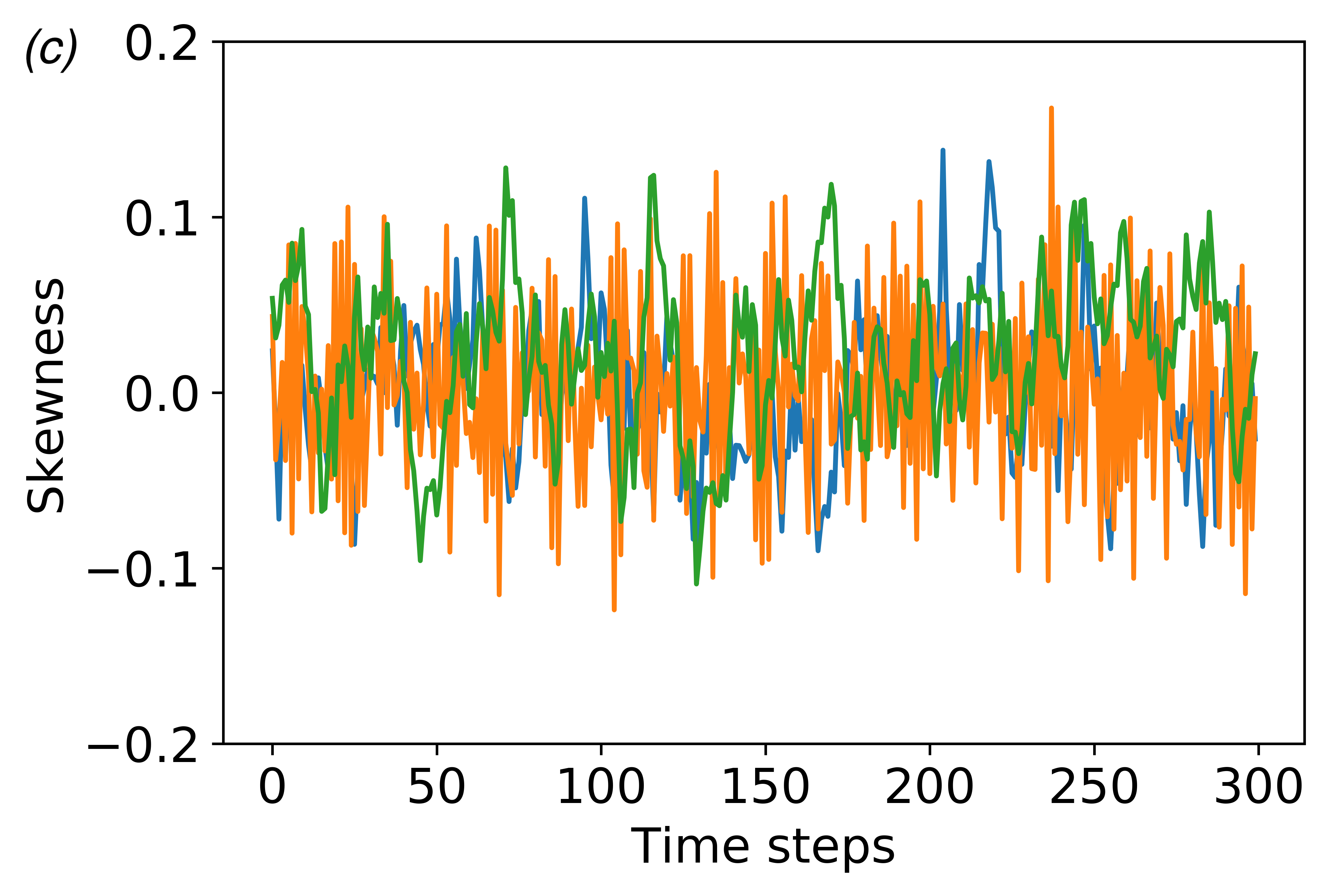}
			\includegraphics[width=0.7\linewidth]{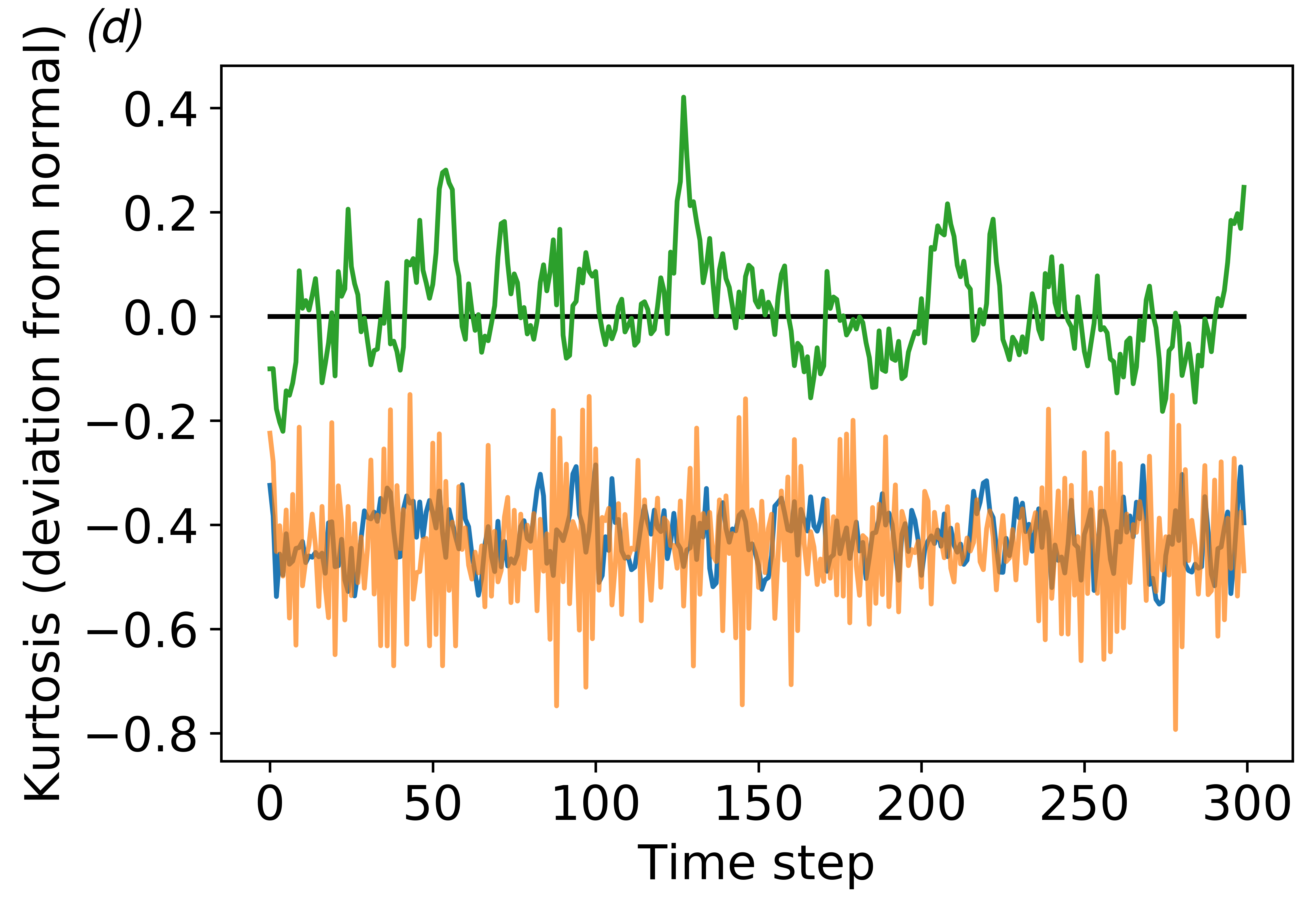}
	\caption{\small{ (a) Distributions for the three different processes: classical process in a confining quartic potential (blue circles), McKean-Vlasov process in a quartic potential (orange triangles), classical Ornstein-Uhlenbeck process in a quadratic potential (green stars); (b) variances of the different processes (same color codes); (c) skewness; and (d) kurtosis. The small skewness reveals the symmetry of the distributions. From the variance and the kurtosis, we note that, despite the flattening of the distributions due to the kernel size, the specificity of each case is still clearly visible.}}
	\label{fig:moments}
\end{figure}

In Fig. \ref{fig:moments}, we represent a logarithmic plot of the final distribution of $\mathcal{N} =3000 $ simultaneous trajectories for three different cases: (i) classical process in a confining quartic potential, (ii) McKean-Vlasov process in a quartic potential, and (iii) classical Ornstein-Uhlenbeck process in a quadratic potential.
We also show the fitting of the distribution with two models: the Ornstein-Uhlenbeck final distribution is fitted with a Gaussian distribution, while the other two cases are fitted with a model of the form $P = p_1 e^{p_2x^2 + p_3x^4}$, which is a classical Maxwell-Boltzmann distribution in a quartic potential. We see that the McKean-Vlasov final distribution also belongs to this category, which means that the Bohm potential in this case takes the form of a quartic-type potential.
We also represent the first moments of the distributions in order to obtain more detailed information. We note that the quantum and classical cases in a quartic potential (orange and blue lines)  differ mostly by their respective variances, whereas their kurtosis (describing their tailedness) are similar, departing noticeably from the Gaussian value. The skewnesses, as expected, are zero since all distributions are symmetrical.

\begin{figure}[ht]
	\vspace{5mm}
	\centering
	\includegraphics[width=0.7\linewidth]{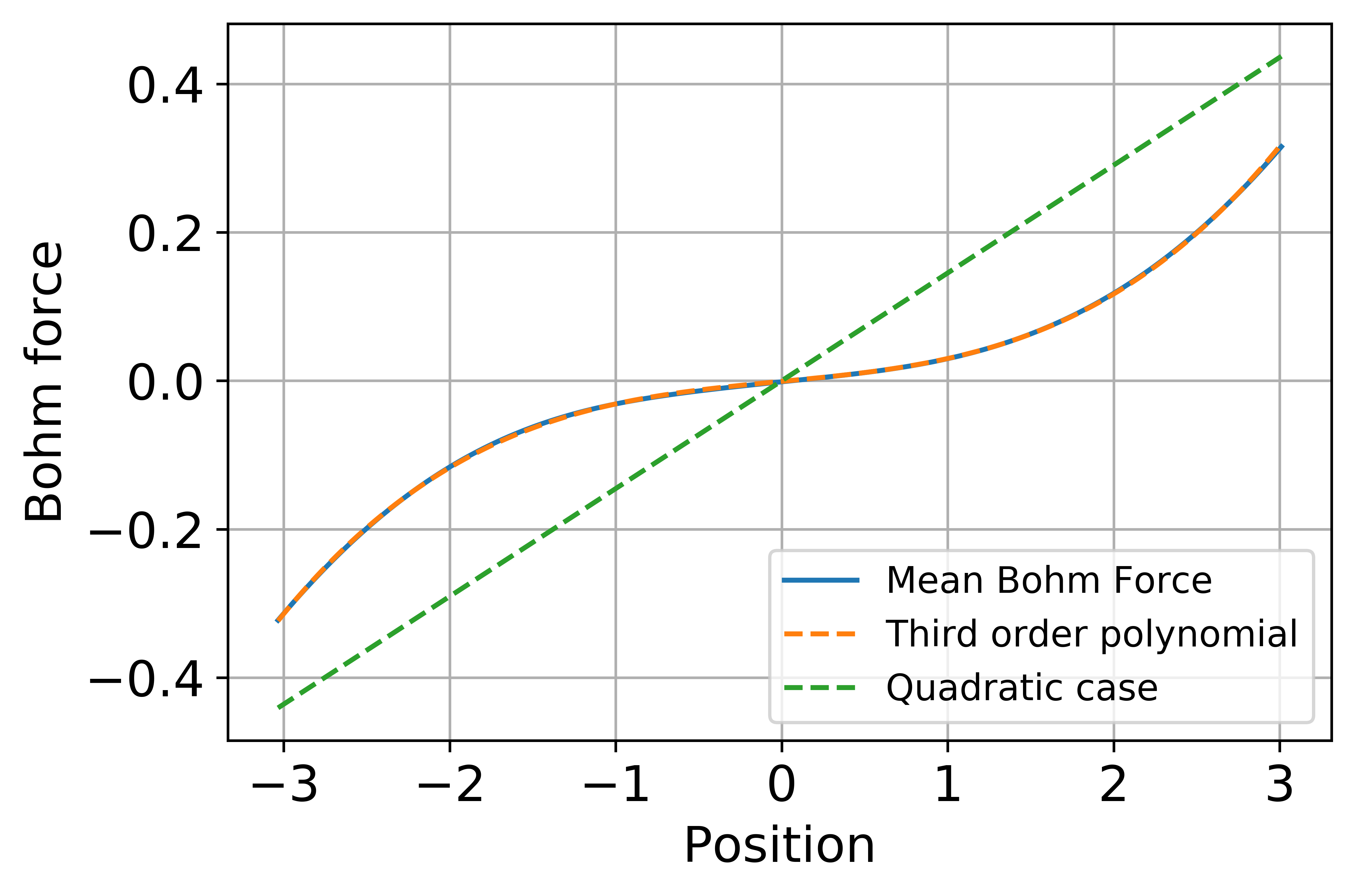}
	\caption{\small{
		Mean Bohm force in the case of a quartic external potential, with a third order polynomial fit. We also show the quadratic limit, which is simply $\sim x/S^2$.
	}}
	\label{fig:DistFit}
\end{figure}
\vspace{5mm}

In Fig. \ref{fig:DistFit} we represent the Bohm force in the quartic case, along with a third order polynomial fit and the Gaussian limit.
We see that the agreement of the fit is good, showing that in this case again, the Bohm potential takes a form similar to the external potential.

\section{Experimental setup and calibration}
\label{APP:setup}

\begin{figure}[ht]
	\centering
	\includegraphics[width=0.9\linewidth]{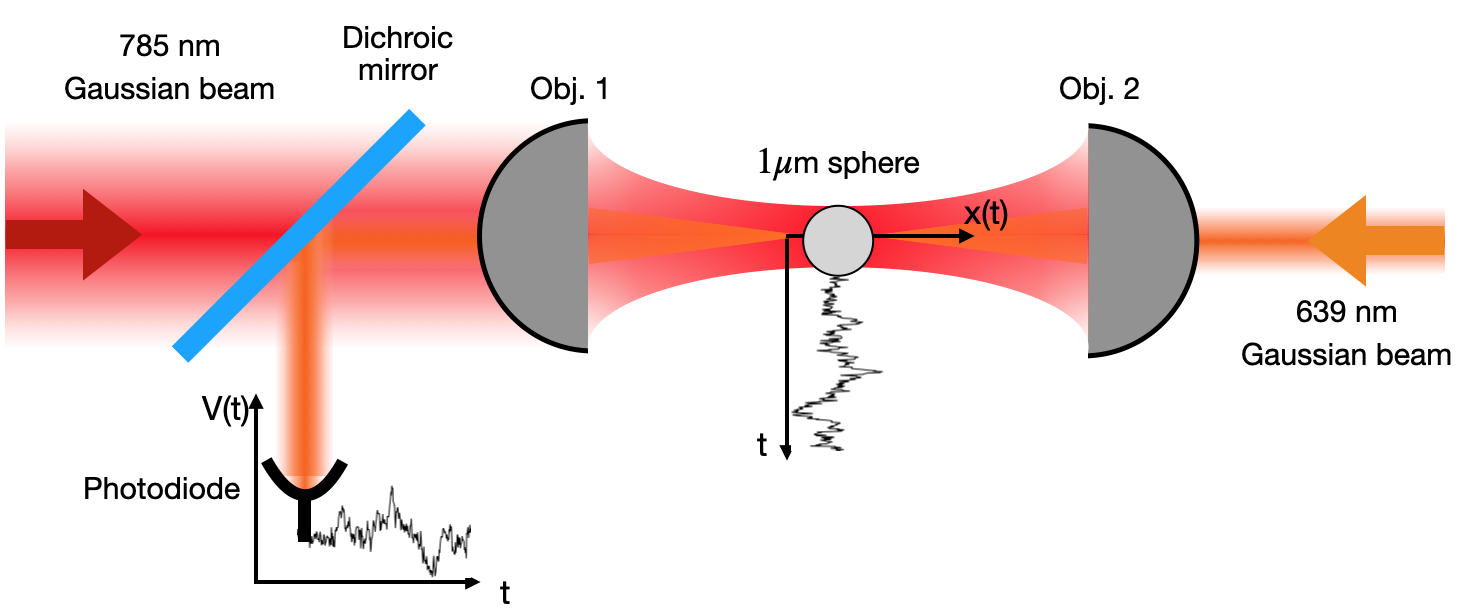}
	\caption{\small{
			Schematic view of the experimental setup: a $1~\mu$m dielectric bead immersed in water is harmonically trapped in the waist of a focused 785 nm laser Gaussian beam. Its position along the optical axis $x$ is recorded with a 639 nm low power probe laser. Both beam are separated using a dichroic mirror and the probe signal is recorded using a photodiode. The intensity signal is linearly dependent on the bead displacement and can be calibrated in order to obtain $x(t)$
	}}
	\label{fig:schema-app}
\end{figure}

Our experiment, schematized in Fig. \ref{fig:schema-app}, consists in trapping a single Brownian object in the harmonic potential created by a focused laser beam.
A linearly polarized Gaussian beam (CW 785 nm diode) is focused by a water immersion objective (Nikon Plan-Apo VC, $60\times$ Numerical Aperture $1.20$ Water Immersion) into a fluidic cell of $120~\mu$m thickness filled with deionized water with a monodispersed suspension of polystyrene microspheres (Thermoscientific Fluoro-Max, radius $500$ nm).
We make sure that only one single bead is trapped at the waist of the focused beam, using an interferometric scattering microscopy system (not shown on the figure) \cite{Lindfors2004}.

The position of the bead is recorded using a low-power counter-propagating laser beam (639 nm diode), focused on the bead using a second objective (Nikon Plan-fluo ELWD 60x0.70). The light scattered by the bead is recollected and send to a photodiode (Thorlabs Det10A).
The signal recorded (in V/s) is send to a low noise amplifier (SR560) and then acquired by an analog-to-digital card (NI PCI-6251). The signal is filtered through a $0.3$ Hz high-pass filter at 6 dB/oct to remove the DC component and through a $100$ kHz low-pas filter at 6 dB/oct to prevent from aliasing.
The position of the bead along the optical axis is, for small enough displacements linear with the scattered intensity.
Furthermore, we work in the linear regime of our photodiodes so that the signal remains linear with the intensity. Finally, the resulting voltage trace is also linear with the instantaneous position $x(t)$ of the trapped bead.

\begin{figure}[ht]
	\centering
	\includegraphics[width=0.9\linewidth]{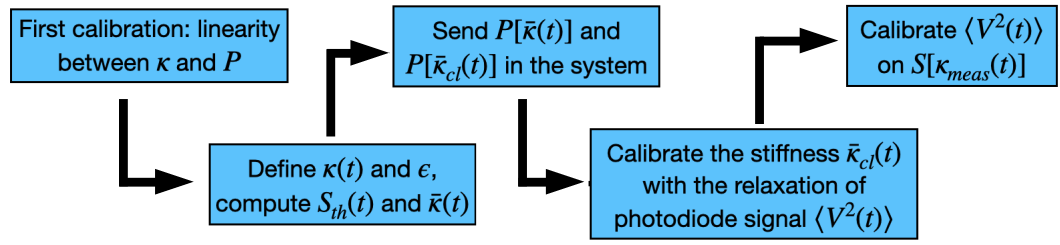}
	\caption{\small{
			Schematic sequence of the different steps of our experimental calibration procedure. $P$ is the trapping laser power, $\kappa$ the stiffness of the harmonic potential, $\bar{\kappa}$ is the modified stiffness that includes quantum effects through $\bar{\kappa}(t) = \kappa(t) - \hbar_{arb}^2/4mS(t)^2$. $\bar{\kappa}_{cl}$ is the step-like protocol connecting the same initial and final stiffnesses, $V(t)$ is the voltage signal of the photodiode recording the beads position along the optical axis $x(t)$, $\kappa_{meas}$ is the stiffness obtained by the relaxation calibration and $S[\kappa_{meas}]$ is the solution of the variance ODE for $\kappa(t) = \kappa_{meas}(t)$.
	}}
	\label{fig:calib}
\end{figure}

In our experimental implementation, the optical potential created by the focused laser beam is locally harmonic.
The stiffness of the harmonic potential $\kappa(t)$ is proportional to the laser power $P(t)$ and can be controlled by the experimentalist.
Our experimental method and calibration are based on the theoretical results obtained in the harmonic and Gaussian case, mainly the relation between the stiffness $\kappa(t)$ and the variance $S(t)$ given by  Eq. (\ref{Eq:VarGaussian}).
It makes it possible to realize the McKean-Vlasov process using one single trajectory and to use this system to probe out-of-equilibrium states, with a given protocol $\kappa(t)$.
The method is the following: first a protocol $\kappa(t)$ is defined, a value of the parameter $\epsilon$ is chosen and  is transferred to an arbitrary Planck constant $\hbar_{arb}^2 = \frac{2 m (k_B T \epsilon)^2}{\kappa_i}$. Then the variance ODE is solved for this protocol and the modified stiffness $\bar{\kappa}(t) = \kappa(t) - \hbar_{arb}^2/4mS^2(t)$ is injected as a laser intensity protocol.
The different steps of the procedure including the calibration are summarized Fig. \ref{fig:calib}.

This procedure however relies on a precise calibration of the system: in order to use the variance differential equation, we need to know with the best possible precision the stiffness $\kappa(t)$ at play in the trap. In this section, we detail our method.

\vspace{5mm}

\begin{figure}[h!]
		\includegraphics[width=0.7\linewidth]{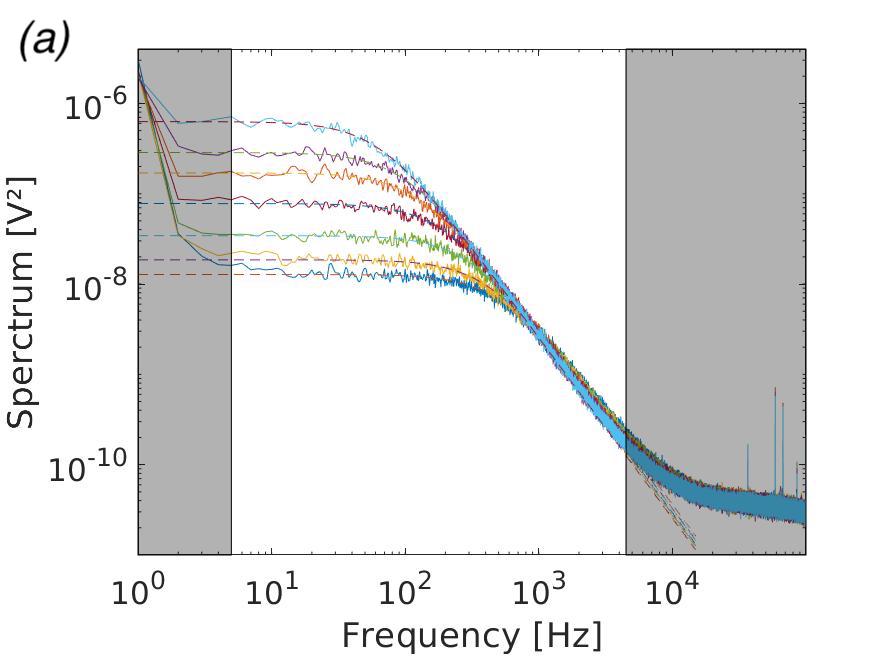}
		\includegraphics[width=0.7\linewidth]{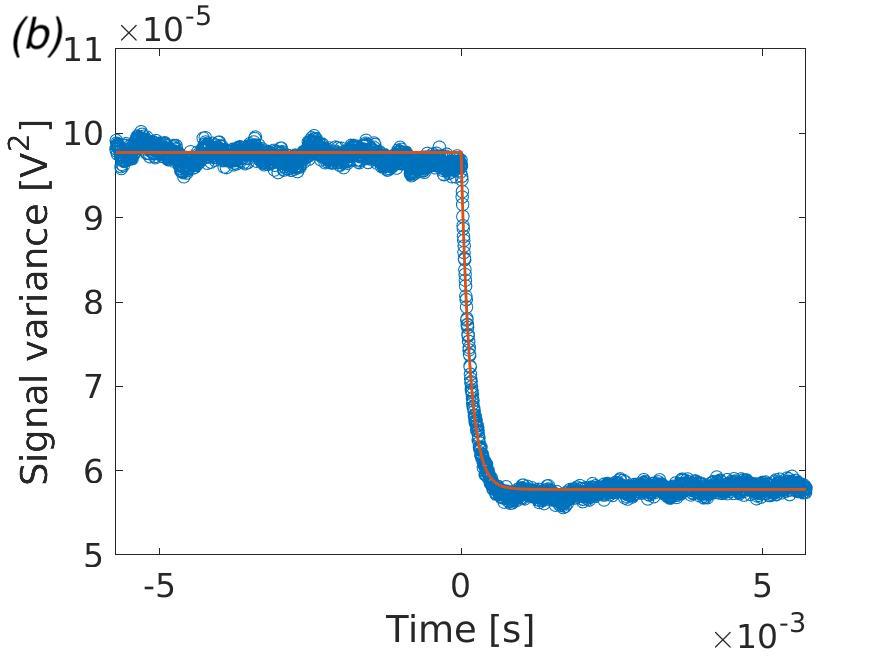}
		\includegraphics[width=0.7\linewidth]{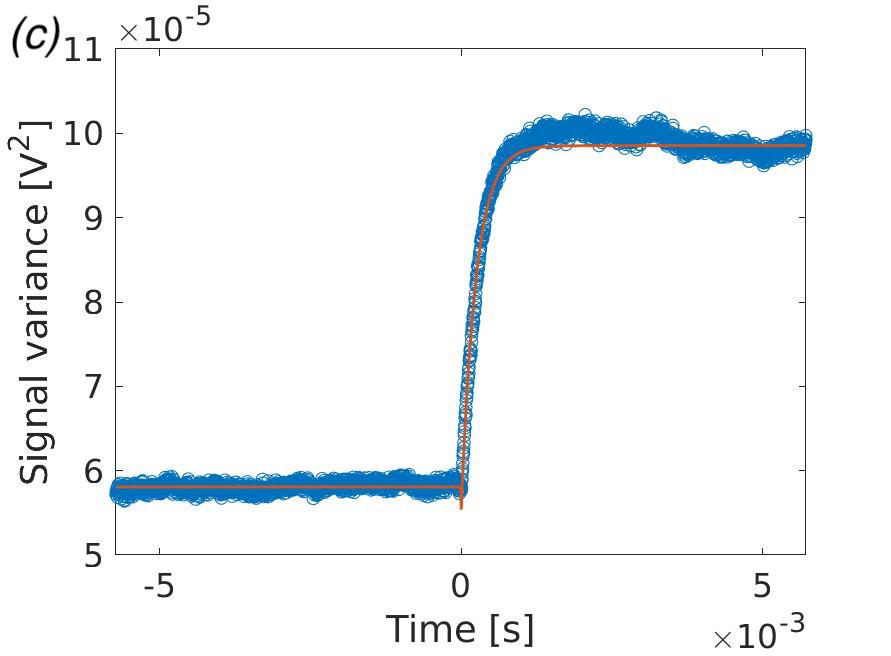}
	\caption{\small{
		(a) Power spectral densities of the measured photodiode voltages for different trapping stiffnesses with Lorentzian fit, (b) exponential decay fit of the "up" step from $\kappa_i$ to $\kappa_f$ ($\kappa_f > \kappa_i$), allowing us to obtain a measure of $\kappa_f$, (c) exponential decay fit of the "down" step from $\kappa_f$ to $\kappa_i$, allowing us to obtain a measure of $\kappa_i$.
	}}
	\label{fig:APPcalib}
\end{figure}

In order to predict the stiffness in the trap, we first calibrate the linear relation between the trapping laser power $P(t)$ and $\kappa(t)$.
We use the power spectral density (PSD) method \cite{berg_sorensen}. The Ornstein-Uhlenbeck process describes the Brownian motion in the trap
\begin{equation*}
dx_t = - (\kappa/\gamma) x_t dt + \sqrt{2 kb T / \gamma}dW_t  ,
\end{equation*}
where $\gamma$ is Stokes coefficient given by $\gamma = 6 \pi \eta R$ where $\eta$ is water viscosity and $R$ the beads radius. This process can be spectrally analysed with the position PSD:
\begin{equation}
 	S_{x}(f) = \frac{D}{\pi^2(f_c^2 + f^2)} .
 	\label{psd_eq}
 \end{equation}
where the roll-off frequency $f_c = \kappa/ 2 \pi \gamma$ separates the high frequency regime $S_x(f)\sim{D}/{\pi^2f^2}$ of free Brownian motion from the low frequency trapping regime $S_{x}(f) \sim {D}/{\pi^2 f_c^2} = 4 k_B T \gamma / \kappa $.
By recording a trajectory with a certain laser power, one can obtain the stiffness $\kappa$ from the roll-off $f_c$, by a Lorentzian fit of the spectrum.
In Fig. \ref{fig:APPcalib} (left) we represent the PSD and fit for different trapping strength that gives the linear relation between $\kappa$ and the laser power.
It is then possible to send a designed protocol of stiffness $\kappa(t)$ by inverting the relation.

\vspace{3mm}

In order to build an ensemble of synchronised trajectories experiencing a defined protocol, we rely on the ergodic hypothesis.
From one long trajectory experiencing a series of protocols, we build an ensemble of $N_{exp} \approx 2 \times 10^4$ trajectories.
We start by defining a step-like protocol where $\kappa(t)$ goes abruptly from $\kappa_i$ to $\kappa_f$ and send it as $P(t)$ to the trapping laser.
From the obtained ensemble of trajectories experiencing a transient relaxation, we extract the photodiode signal variance $\langle V^2(t) \rangle$ that follows an exponential decay (solution of the classical Fokker-Planck equation).
This decay $\sim e^{-\kappa_f t / \gamma}$ is fully characterising the final stiffness.
With an exponential fit of both the "up" and "down" stiffness steps, we recover a measurement of the stiffness performed in the time-domain.
This allows us to measure, during the experiment the actual stiffness at play that can depart slightly from the expected value, due to small drifts or to fit errors of the Lorentzian \cite{berg_sorensen}.
Since we double each McKean-Vlasov experiment with an equivalent classical step, we can perform this dynamical calibration for each experiment.

After the first step-like protocol experiment, we define a value of $\epsilon$ and perform both the quantum and the classical analog experiments.
The dynamical calibration gives the values of $\kappa_i$ and $\kappa_f$, that yield an $\epsilon$ that can slightly differ from the predicted value. These values correspond to the parameters needed for the analytical results.

\begin{figure}[h!]
			\includegraphics[width=0.7\linewidth]{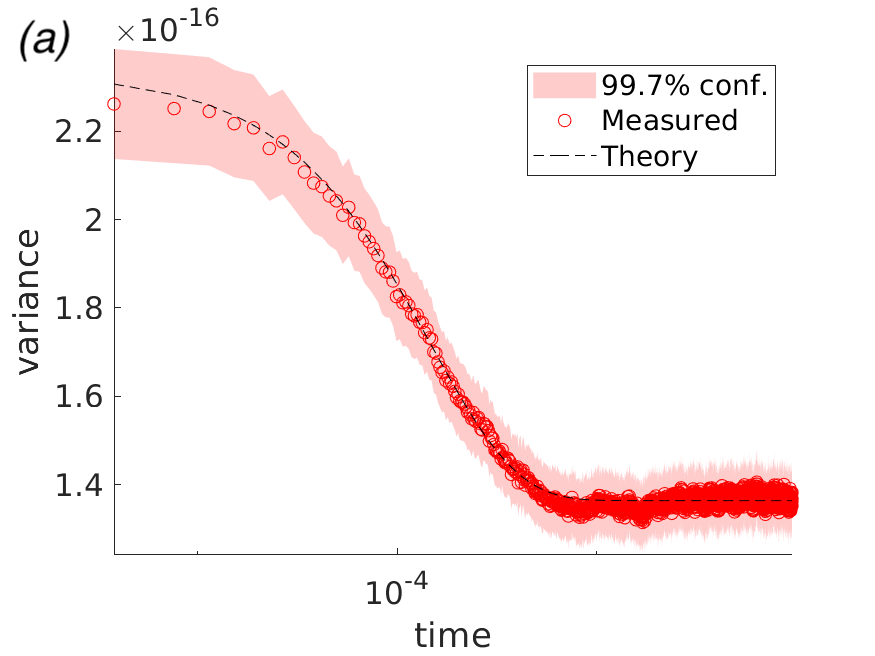}
			\includegraphics[width=0.7\linewidth]{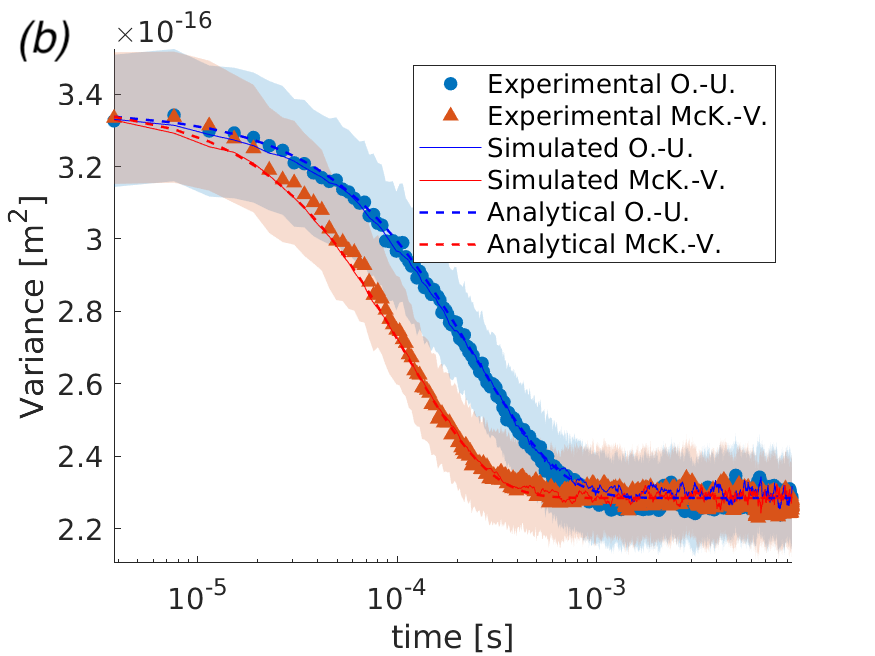}
			\includegraphics[width=0.7\linewidth]{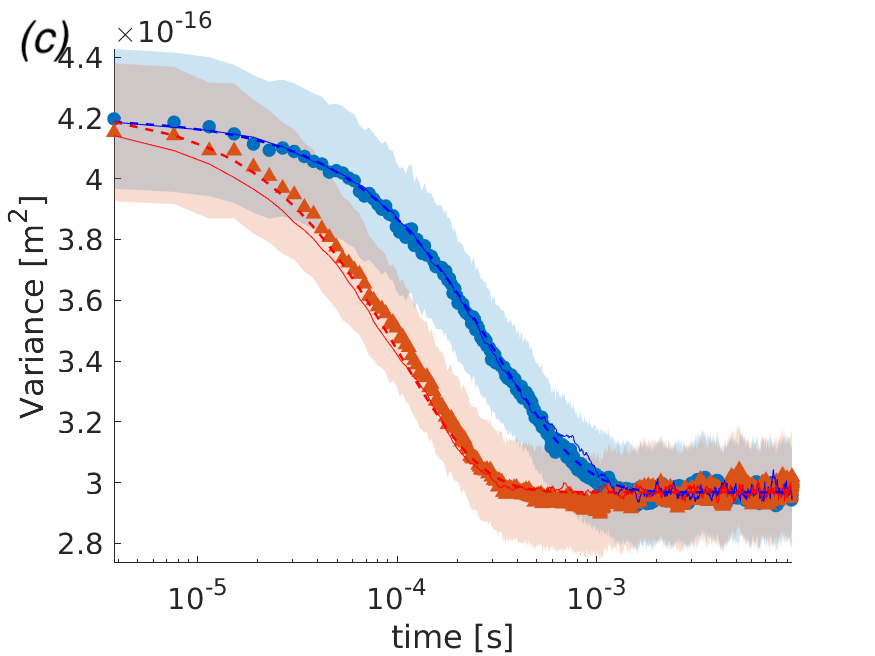}
			\includegraphics[width=0.7\linewidth]{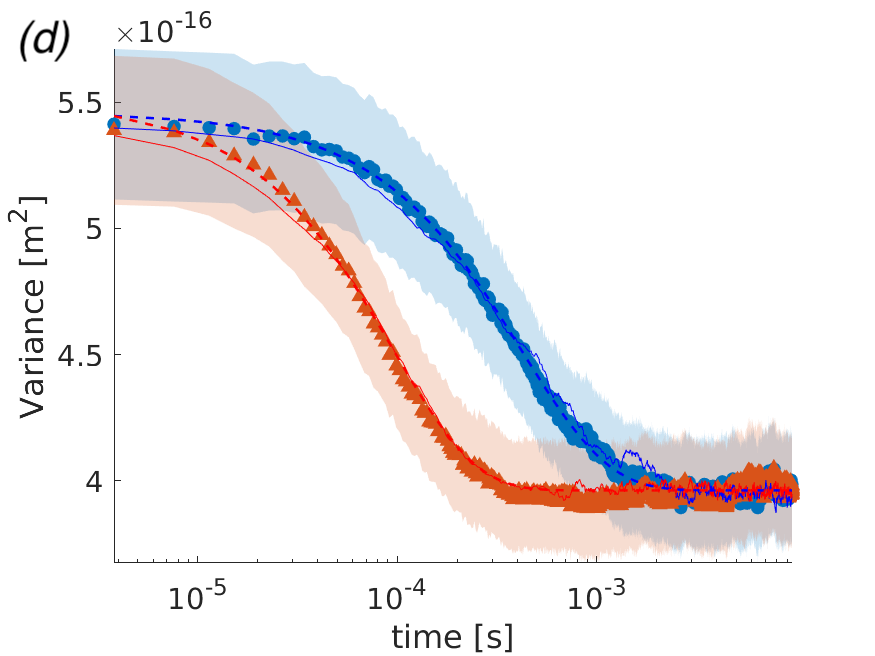}
	\caption{\small{Experimental results: we show the evolution of the variance of an ensemble of over $2 \times 10^4$ trajectories during an out-of-equilibrium transition between two harmonic confinements with different stiffness. The corresponding values of the $\epsilon$ parameter governing the strength of the Bohm force are $\epsilon = 0$ (a); $\epsilon = 1.0558$ (b);  $\epsilon = 1.4089$ (c); and  $\epsilon = 1.801$ (d). For all cases, we show the result of the McKean-Vlasov dynamics (orange triangles) as well as a classical equivalent Ornstein-Uhlenbeck dynamic (blue circles) experiencing a transition between the same initial and final distributions.
	On each curve, we superimpose the result of a numerical simulation performed by measuring at each time-step the ensemble variance of $N = 2 \times 10^4$ simultaneous trajectories and reinjecting it in the next time-step (respectively red and blue thin solid lines).
	We also show the result of the variance differential equation (respectively red and blue thick dashed lines). }}
	\label{fig:APPExpResults}
\end{figure}

Furthermore, the variance $S(t)$ and the stiffness $\kappa(t)$ are unambiguously connected by the variance differential equation $ \frac{d S(t)}{dt} = - \frac{2\bar{\kappa(t)}}{\gamma} S(t) + \frac{2k_B T}{\gamma} $. Hence, once we know the stiffness, we can compute $S(t)$ and can then calibrate our measured voltage variance $\langle V^2(t) \rangle $ to $S(t)$.
We fit the transformation by a linear relationship $S(t) = \alpha \langle V^2(t) \rangle  + \beta$ which implies that the position transforms according to $x_t = \sqrt{\alpha} V(t) + \sqrt{\beta} \mathcal{N}(0,1)$
where the first term represents the linear response of the photodiode and the second the sum of all experimental noises (that we approximate as a resulting white noise).
This method allows a precise calibration of the variance, as seen on Fig. \ref{fig:APPExpResults}, which is our only observable here. This method gives the position itself only up to the noise therm, which is several order of magnitude smaller than the $\alpha$ term.

The error on the experimental variance essentially comes from three main sources.
One is the error on the experimental parameters such as the temperature or the radius of the trapped bead, through the viscous drag coefficient $\gamma $.
It is dominated by the $2.8\%$ uncertainties on the beads radius $R$ that result in a similar error on $\gamma = 6 \pi \eta R$ where $\eta$ is the water viscosity.
Other sources of errors (temperature) are also taken into account but their final influence is not significant. Temperature in particular is controlled with a precision better than $1$ K.
The error on the radius is simply taken into account by carrying the whole analysis with the two "worst" values of radius, yielding an error $\delta_{param} \approx 3\%$ between the two extreme results.
The second source of errors is the statistical reliability of an estimator of the variance on an ensemble of finite size. It is obtained following the $\chi^2$ test on $N-1$ degrees of freedom, where $N$ is the size of the ensemble.
We carry the test with $3 \sigma = 99.7\%$ confidence interval giving $ \delta_{\chi^2}$.
The third source is the error arising from the fitting procedure in the calibration of the decay to obtain the stiffness and to calibrate the variance from \si{\squared\volt} to \si{\square\meter} giving $\delta_{fit}$.
The obtained variance is then defined up to:
\begin{equation}
S_{exp} = \langle x^2(t) \rangle \pm \left( \delta_{\chi^2}(t) + \delta_{param}(t)  + \delta_{fit}(t)\right).
\end{equation}
These different error sources give the colored patch shown on each plot of Fig. \ref{fig:APPExpResults}.

\bibliography{biblioMcKV}

\end{document}